\documentclass{nature}

\usepackage{graphicx}
\usepackage{dcolumn}
\usepackage{bm}
\usepackage{ulem}
\usepackage{amssymb}
\usepackage{multirow}
\usepackage{lineno}
\usepackage{amsmath}
\usepackage[colorlinks=true,citecolor=blue]{hyperref}

\makeatletter
\let\saved@includegraphics\includegraphics
\AtBeginDocument{\let\includegraphics\saved@includegraphics}
\renewenvironment*{figure}{\@float{figure}}{\end@float}
\makeatother

\title{Interplay between Kitaev interaction and single ion anisotropy in ferromagnetic CrI$_3$ and CrGeTe$_3$ monolayers}
\author{Changsong Xu$^{1}$, Junsheng Feng$^{2,3}$, Hongjun Xiang$^{2,4,*}$, \& L. Bellaiche$^{1,*}$}

\begin{document}

\maketitle


\begin{affiliations}
 \item Physics Department and Institute for Nanoscience and Engineering, University of Arkansas, Fayetteville, Arkansas 72701, USA
 \item Key Laboratory of Computational Physical Sciences (Ministry of Education), State Key Laboratory of Surface Physics, and Department of Physics, Fudan University, Shanghai, 200433, China
 \item School of Electronic and Information Engineering, Hefei Normal University, Hefei 230601, P. R. China
 \item Collaborative Innovation Center of Advanced Microstructures, Nanjing 210093, P. R. China

\end{affiliations}

\begin{abstract}
  Magnetic anisotropy is crucially important for the stabilization of two-dimensional (2D) magnetism, which is rare in nature but highly desirable in spintronics and for advancing fundamental knowledge. Recent works on CrI$_3$ and CrGeTe$_3$ monolayers not only led to observations of the long-time-sought 2D ferromagnetism, but also revealed distinct magnetic anisotropy in the two systems, namely Ising behavior for CrI$_3$ versus Heisenberg behavior for CrGeTe$_3$. Such magnetic difference strongly contrasts with structural and electronic similarities of these two materials, and understanding it at a microscopic scale should be of large benefits.  Here, first-principles calculations are performed and analyzed to develop a simple Hamiltonian, to investigate magnetic anisotropy of CrI$_3$ and CrGeTe$_3$ monolayers. The anisotropic exchange coupling in both systems is surprisingly determined to be of Kitaev-type. Moreover, the interplay between this Kitaev interaction and single ion anisotropy (SIA) is found to naturally explain the different magnetic behaviors of CrI$_3$ and CrGeTe$_3$.  Finally, both the Kitaev interaction and SIA are further found to be induced by spin-orbit coupling of the heavy ligands (I of CrI$_3$ or Te of CrGeTe$_3$) rather than the commonly believed 3d magnetic Cr ions.
\end{abstract}

\noindent
{\bf INTRODUCTION}\\
Two-dimensional (2D) magnetic materials are receiving a lot of attention, due, e.g., to the search for long range ferromagnetism (FM)\cite{griffiths1964peierls,mermin1966absence}, which can facilitate various applications from sensing to data storage\cite{soumyan2016emergent,miao20182d}. According to Mermin and Wagner¡¯s theorem\cite{mermin1966absence}, however strong the short-range isotropic couplings are, the realization of 2D magnetism relies on magnetic anisotropy, as a result of spin-orbit coupling (SOC). The requirement of strong magnetic anisotropy in low dimensional systems therefore explains  the rareness of 2D FM materials.

The recent observation of ferromagnetism in monolayers made of CrI$_3$ and CrGeTe$_3$ \cite{huang2017layer,gong2017discovery,xing2017electric} therefore opens a new chapter in the  field of 2D materials. The chromium in both compounds share the same valence state of Cr$^{3+}$, with the 3$d^3$ configuration and S = $\frac{3}{2}$\cite{lado2017origin,li2014crxte,xing2017electric,carteaux1995crystallographic}. Ferromagnetism arises there from the super exchange between nearest neighbor Cr ions, that are linked by I or Te ligands through nearly 90$^\circ$ angles\cite{carteaux1995crystallographic,mcguire2015coupling}. CrI$_3$ has been demonstrated to be well described by the Ising behavior\cite{huang2017layer,samarth2017condensed,griffiths1964peierls}, for which the spins can point up and down along the out-of-plane z-direction. In contrast, the magnetic anisotropy of CrGeTe$_3$ was determined to be consistent with the Heisenberg behavior\cite{samarth2017condensed,xing2017electric,mermin1966absence},  for which the spins can freely rotate and adopt any direction in the three-dimensional space. Interestingly, structural and electronic similarities between these two compounds strongly contrast with their difference in magnetic behaviors, which implies subtle origins for their magnetic anisotropy. A recent theoretical work adopted the XXZ model, for which the exchange coupling is identical between the in-plane x- and y-directions but different along the out-of-plane z-direction,  to explain the   out-of-plane magnetization of CrI$_3$\cite{lado2017origin}. However, there is no definite proof that the XXZ model is accurate enough to describe the magnetic anisotropy of CrI$_3$, and there is  a  current paucity of knowledge for the mechanism responsible for the magnetic anisotropy of CrGeTe$_3$. Hence, a thorough microscopic understanding of the difference between the Ising behavior of CrI$_3$ and the Heisenberg behavior of CrGeTe$_3$ is highly desired.

In particular, it is tempting to investigate if the Kitaev interaction\cite{kitaev2006anyons}, which is a  specific  anisotropic exchange coupling, can also be significant in CrI$_3$ and CrGeTe$_3$. This temptation is mainly based on the fact that these two materials adopt a honeycomb lattice and edge-sharing octahedra, exactly as the layered Na$_2$IrO$_3$ and $\alpha$-RuCl$_3$ compounds which exhibit magnetic behaviors that are close to spin liquids\cite{jackeli2009mott}  --as a result of significant Kitaev interactions.
Interestingly, finding  finite Kitaev interaction in Cr-3$d$ based CrI$_3$ and CrGeTe$_3$ compounds would  enlarge the types of systems possessing such interaction, not only from $4d$ or $5d$ to $3d$ transition-metal-based insulators, but also from S = 1/2 to S = 3/2 systems. Such broadening is in-line with recent theoretical predictions of Kitaev interaction in $d^7$ or $3d$ systems\cite{liu2018pseudospin,sano2018kitaev}.

The main goal of this manuscript is to report results of first-principles calculations, along with the concomitant development of a simple but predictive Hamiltonian, to demonstrate that: (i) significant Kitaev interaction does exist in  CrI$_3$ and CrGeTe$_3$ (which also  invalidates the XXZ model in these two compounds); and (ii) the different interplay between this Kitaev interaction and the single ion anisotropy (SIA) naturally explains the observed magnetic anisotropy in these 2D ferromagnetic materials.  Another surprising result is that the Kitaev and SIA anisotropies are both dominantly induced by the SOC of the heavy ligand elements rather than the $3d$ element Cr.

\noindent
{\bf RESULTS}\\
To precisely describe the magnetic anisotropy and explore differences between CrI$_3$ and CrGeTe$_3$, we consider a Hamiltonian containing both exchange coupling, $\mathcal{H}_{ex}$, and SIA, $\mathcal{H}_{si}$, terms:
\begin{linenomath*}
\begin{equation}
  \mathcal{H} = \mathcal{H}_{ex} + \mathcal{H}_{si}
    = \frac{1}{2} \sum_{i,j} \bm{{\rm S}}_i {\cdot} \mathcal{J}_{ij} {\cdot} \bm{{\rm S}}_j + \sum_{i} \bm{{\rm S}}_i {\cdot} \mathcal{A}_{ii} {\cdot} \bm{{\rm S}}_i
\end{equation}
\end{linenomath*}
where $\mathcal{J}_{ij}$ and $\mathcal{A}_{ii}$ are 3$\times$3 matrices gathering exchange and SIA parameters, respectively. The sum over $i$ in Eq. (1) runs over all Cr sites, while the sums over $i,j$ run over all nearest neighbor Cr pairs(note that the anisotropy in exchange coupling between more distant Cr neighbors is at least an order smaller and is thus negligible, see details in Supplementary Materials (SM)\cite{Note1}). DFT calculations are  performed on CrI$_3$ and CrGeTe$_3$ monolayers to extract the components of $\mathcal{J}$ and $\mathcal{A}$ using a precise four-states method\cite{xiang2013magnetic} (see SM for details\cite{Note1}). Note that (i) all the results shown below are based on the use of an effective Hubbard $U$ = 0.5 eV, unless  stated (see the effects of the choice of other $U$'s and details of method in SM\cite{Note1}); and (ii) the Dzyaloshinski-Moriya (DM) interaction is absent in our studied systems because of the existence of an inversion center between nearest neighbor Cr ions\cite{moriya1960anisotropic}.

\begin{figure}
\includegraphics[width=10cm]{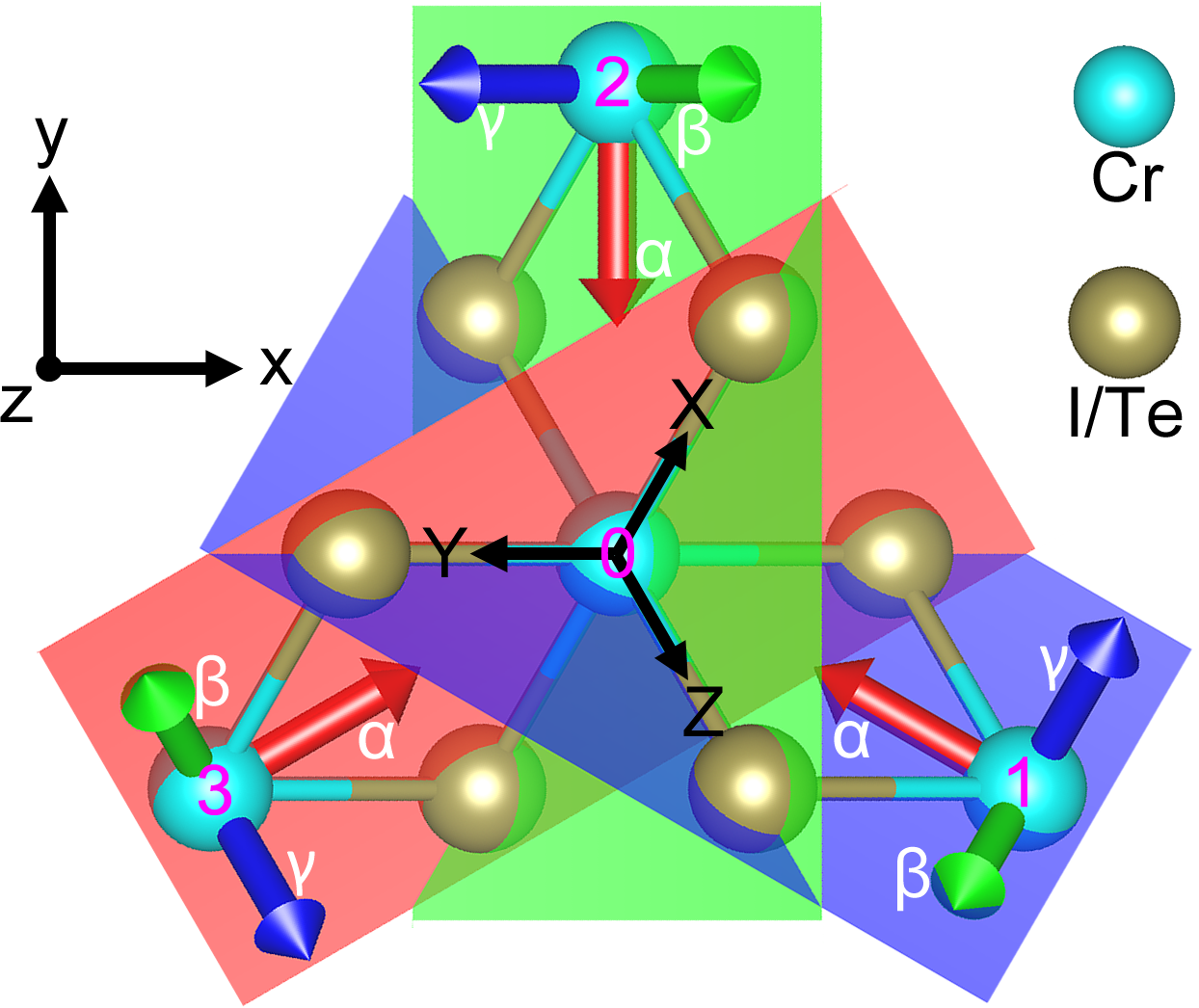}\centering%
\caption{Schematization of the CrI$_3$ and CrGeTe$_3$ structures, as well as the different coordinate systems indicated in the text. Note that Ge of CrGeTe$_3$ is not shown for simplicity.}
\end{figure}

Let us first focus on the exchange coupling $\mathcal{J}$ matrix for CrI$_3$ and CrGeTe$_3$. The $\mathcal{J}$ matrix is expressed in the \{xyz\} basis, for which the x-y plane is the film plane while the z-axis is the out-of-plane direction of the film.  We choose the Cr0-Cr1 pair (see Fig. 1) to calculate the exchange coupling parameters, from which the parameters for Cr0-Cr2 and Cr0-Cr3 pairs can be deduced via three-fold rotational symmetry. It is numerically found that this matrix is symmetric, i.e., $J_{xy}$=$J_{yx}$, $J_{yz}$=$J_{zy}$ and $J_{xz}$=$J_{zx}$, which is consistent with the fact that there is no DM interaction in our investigated compounds.  As shown in Table 1, $J_{xx}$, $J_{yy}$ and $J_{zz}$ of CrI$_3$ possess quite different values of -2.29, -1.93 and -2.23 meV, respectively, while the off-diagonal elements of $\mathcal{J}$ in the \{xyz\} basis are smaller but non-negligible. Such results  contrast with the  XXZ model adopted in Ref.\cite{lado2017origin} and that assumes that  (i) $J_{xx} = J_{yy}  \neq J_{zz}$; and (ii) $J_{xy}$, $J_{yz}$ and $J_{xz}$ can all be neglected. Different schemes and strategies, such as changing the value of $U$, using experimental structures rather than the computationally optimized ones and even replacing other Cr ions by nonmagnetic Al, are used to check their influence on $J_{xx}$, $J_{yy}$ and $J_{zz}$ of CrI$_3$. It is numerically found that they all qualitatively give the same results (as detailed in SM\cite{Note1}) in the sense that the aforementioned assumption (i) of the XXZ  model providing equality between $J_{xx}$ and $J_{yy}$ is not satisfied, which automatically implies that such latter model is not accurate enough to precisely describe  magnetic anisotropy in CrI$_3$ and CrGeTe$_3$ systems.

\begin{table}\centering
  \caption{Matrix components of $\mathcal{J}$, as well as the $J$ and $K$ parameters, of the Cr0-Cr1 pair, and the SIA coefficient $A_{zz}$ for the Cr0 ion, as given by DFT calculations with $S$ = $\frac{3}{2}$.
  Corresponding coefficients for the Cr0-Cr2 and Cr0-Cr3 pairs can be deduced via three-fold rotational symmetry. Note that the global coordinate system \{xyz\} and the local coordinate system \{$\alpha\beta\gamma$\} of each Cr-Cr pair are shown in Fig. 1. The units of the parameters indicated here is meV.}
  \begin{tabular}{>{\hfil}p{50pt}<{\hfil}>{\hfil}p{29pt}<{\hfil}>{\hfil}p{29pt}<{\hfil}>{\hfil}p{29pt}<{\hfil}>{\hfil}p{29pt}<{\hfil}>{\hfil}p{29pt}<{\hfil}>{\hfil}p{29pt}<{\hfil}}
  \hline\hline
   & $J_{xx}$ & $J_{yy}$ & $J_{zz}$ & $J_{xy}$ & $J_{yz}$ & $J_{xz}$ \\
  \hline
   CrI$_3$    & -2.29 & -1.93 & -2.23 & 0.30 & ~0.29 & ~0.17 \\
   CrGeTe$_3$ & -6.54 & -6.37 & -6.64 & 0.15 & -0.04 & -0.02\\
  \hline
   & $J_{\alpha}$ & $J_{\beta}$ & $J_{\gamma}$ & $J$ &$K$ & $A_{zz}$ \\
  \hline
   CrI$_3$    & -2.46 & -2.41 & -1.59 & -2.44 & 0.85  & -0.26\\
   CrGeTe$_3$ & -6.65 & -6.63 & -6.28 & -6.64 & 0.36  & ~0.25\\
  \hline
  \end{tabular}
\end{table}

\begin{table}
\centering
  \caption{Parameters of Eq. (5) and $\Delta{\mathcal{E}}$ from both Eq. (5) and DFT for CrI$_3$ and CrGeTe$_3$. $\Delta{\mathcal{E}}$ defines the energy difference between the energy for an out-of-plane  magnetization and the averaged energy of ferromagnetic states having in-plane magnetization. }
  \begin{tabular}{>{\hfil}p{50pt}<{\hfil}>{\hfil}p{32pt}<{\hfil}>{\hfil}p{32pt}<{\hfil}>{\hfil}p{50pt}<{\hfil}>{\hfil}p{2pt}<{\hfil}>{\hfil}p{33pt}<{\hfil}>{\hfil}p{33pt}<{\hfil}}
  \hline\hline
  \multirow{2}{*}{} & \multicolumn{3}{c}{Parameters (meV)} && \multicolumn{2}{c}{$\Delta{\mathcal{E}}$ (meV/f.u.)}\\
   \cline{2-4} \cline{6-7}
                 & $bK$ & $\frac{2}{3}A_{zz}$ &$bK$+$\frac{2}{3}A_{zz}$ && Eq.(5) & DFT \\
  \hline
   CrI$_3$       & -0.16  & -0.17  & -0.33~ && -1.11 & -0.82\\
   CrGeTe$_3$    & -0.17  & ~0.17  & -0.003 && -0.01 & ~0.02\\
  \hline
  \end{tabular}
\end{table}

The symmetric $\mathcal{J}$ matrix is then diagonalized   to obtain its eigenvalues (to be denoted as $J_{\alpha}$, $J_{\beta}$ and $J_{\gamma}$) and corresponding eigenvectors (to be coined $\alpha$, $\beta$ and $\gamma$) for the Cr0-Cr1 pair. As seen in Table 2, $J_{\alpha}$ and  $J_{\beta}$ are the strongest eigenvalues in  magnitude and are close to each other in CrI$_3$ (-2.46 meV and -2.41 meV, respectively), while  $J_{\gamma}$ is  smaller in magnitude by about 1 meV.  The same hierarchy exists between $J_{\alpha}$, $J_{\beta}$ and $J_{\gamma}$ in  CrGeTe$_3$, but with  $J_{\alpha}$ and $J_{\beta}$ being now stronger in strength (about -6.65 meV), while $J_{\gamma}$ = -6.28 meV is less than 0.4 meV smaller in magnitude than the other two exchange coefficients. As shown in Fig. 1, the $\alpha$-axis points from Cr1 to Cr0 in both systems, and therefore belongs to the x-y plane. On the other hand,  the $\beta$-axis is roughly along the  direction joining the two ligands bridging Cr0 and Cr1, and thus does not belong to the x-y plane. Similarly,   the  $\gamma$-axis, which is perpendicular to both the $\alpha$- and $\beta$-axes, does not lie in the x-y plane.  It is important to realize that the \{$\alpha$$\beta$$\gamma$\} basis diagonalizing the  $\mathcal{J}$ matrix is specific to each considered Cr pairs, unlike the ``global'' \{xyz\} basis. In other words, the \{$\alpha$$\beta$$\gamma$\} basis differs for the  Cr0-Cr1, Cr0-Cr2 and Cr0-Cr3 pairs, as shown in Fig. 1 by means of red, green and blue arrows.

The $\mathcal{H}_{ex}$ exchange coupling Hamiltonian can now be rewritten, gathering these three local orthogonal  \{$\alpha\beta\gamma$\} coordinate bases (one for each Cr pair) and assuming that $J_{\alpha}$ = $J_{\beta}$ (as consistent with the aforementioned calculations\cite{Note2}), as
\begin{linenomath*}
\begin{equation}\label{eq2}
\begin{aligned}
  \mathcal{H}_{ex} = &\frac{1}{2} \sum_{i,j} (  J_{\alpha}S_i^{\alpha}S_j^{\alpha}
                           +J_{\beta}S_i^{\beta}S_j^{\beta}
                           +J_{\gamma}S_i^{\gamma}S_j^{\gamma}  )\\
                        = &\frac{1}{2} \sum_{i,j} ( J\bm{{\rm S}}_i {\cdot} \bm{{\rm S}}_j
                           + KS_i^{\gamma}S_j^{\gamma} )
\end{aligned}
\end{equation}
\end{linenomath*}
where $J$ = $J_{\alpha}$ = $J_{\beta}$ is the isotropic exchange coupling  and $K$ = $J_{\gamma}-J_{\alpha} > 0$ is the so-called Kitaev interaction that characterizes the anisotropic contribution.  Table 1 provides the values of both $J$ and $K$ and, in particular, indicates that the Kitaev interaction can not be neglected in CrI$_3$ and  CrGeTe$_3$.

Let us now investigate the other energy of Eq. (1), that is the SIA, which involves the $\mathcal{A}$ matrix. For that, one needs to go back to the global \{xyz\} basis, since only the  $A_{zz}$ term can be finite by symmetry. It is numerically found that  $A_{zz}$ = -0.26 meV in CrI$_3$, while it adopts a similar magnitude but {\it with a change of sign} in CrGeTe$_3$ (since $A_{zz}$ = 0.25 meV there). Such significant values of $A_{zz}$ (which is of the same order of magnitude than the $K$  Kitaev parameter) implies that SIA is not negligible, which contrasts with the results in Ref.\cite{lado2017origin}.  $\mathcal{H}_{si}$ can thus be simplified as:
\begin{linenomath*}
\begin{equation}
  \mathcal{H}_{si} = \sum_{i} A_{zz}{\rm S}^z_i{\rm S}^z_i
\end{equation}
The total Hamiltonian of Eq. (1) can then be rewritten by combining Eqs. (2) and (3)  as:
\begin{equation}
  \mathcal{H} = \frac{1}{2}\sum_{i,j} ( J\bm{{\rm S}}_i {\cdot} \bm{{\rm S}}_j + KS_i^{\gamma}S_j^{\gamma} )+ \sum_{i} A_{zz}{\rm S}^z_i{\rm S}^z_i
\end{equation}
\end{linenomath*}

This simplified Hamiltonian gathers (i) isotropic exchange coupling from $J$; (ii) anisotropic Kitaev interaction  from $K$ in the different local \{$\alpha\beta\gamma$\} bases; and (iii)  SIA in the global \{xyz\} basis. Let us now try to express the total energy associated with magnetism in an unified coordinate system. Equation (4) shows that the anisotropic part of the exchange energy (arising from $K$) is only related to the projections of spins on the three different $\gamma$-axes (one for each Cr-Cr pair). Due to the fact that these three $\gamma$-axes (to be denoted as $\gamma_1$, $\gamma_2$ and $\gamma_3$, respectively) are normally not perpendicular to each other, we now orthogonalize them using the L{\"o}wdin's symmetric orthogonalization scheme\cite{lowdin1950non}. The resulting orthogonal axes form the \{XYZ\} coordinate system that is shown in Fig. 1.
In this global \{XYZ\} basis, the out-of-plane z-axis of the film is along the [111] direction, and $\gamma_1$, $\gamma_2$ and $\gamma_3$ can be expressed as (1,a,a), (a,1,a) and (a,a,1), where a$\in$[-$\frac{1}{2}$, 1].  As consistent with  the relatively large-in-magnitude and negative value of the isotropic exchange coupling $J$ (see Table 1), ferromagnetic states are considered here. When expressing their spin in the  \{XYZ\} basis, i.e., $\bm{{\rm S}}$($S_{X}$, $S_{Y}$, $S_{Z}$), and considering a magnitude $|\bm{{\rm S}}|=\frac{3}{2}$, it is straightforward to prove that the energy per Cr ion associated with Eq. (4)  can then be rewritten as:
\begin{linenomath*}
\begin{equation}\label{eq3}
\begin{aligned}
   \mathcal{E} = (bK + \frac{2}{3}A_{zz})(S_{X}S_{Y} + S_{Y}S_{Z} + S_{Z}S_{X}) + {\rm C}
\end{aligned}
\end{equation}
\end{linenomath*}
where $b=\frac{a^2+2a}{2a^2+1}$ and C = $\frac{9}{8}(3J+K) + \frac{3}{4}A_{zz}$ are independent of the spin direction (see SM\cite{Note1} for details).  One can also easily demonstrate that the symmetric form of $(S_{X}S_{Y} + S_{Y}S_{Z} + S_{Z}S_{X})$ implies that the magnetization within the x-y plane of the film (for which $S_{X} + S_{Y} + S_{Z} = 0$) is fully isotropic. In other words, any direction of the spin within this x-y plane generates  the same energy.
Note that we further conducted DFT calculations (not shown here) that indeed numerically confirm that such in-plane isotropy is mainly obeyed in CrI$_3$ and CrGeTe$_3$ (the  maximal energetic difference we found  between in-plane directions of spins is 0.006 meV/f.u. and 0.004 meV/f.u. in CrI$_3$ and CrGeTe$_3$, respectively), which attests of the relevance and accuracy of the simple Hamiltonian of Eq. (4) and the resulting energy of Eq. (5).
Note also that, although such isotropy in the x-y plane is in line with the results of Ref.\cite{lado2017origin}, its origin is totally different:  here it arises from the Kitaev interaction and its subsequent frustration, while in Ref.\cite{lado2017origin}, the isotropy in the x-y plane lies in the assumption of the XXZ model (see a detailed comparison between the two models in SM\cite{Note1})

It is also worthwhile to emphasize that Ref.\cite{lado2017origin} assumed that SIA is negligible small, while according to Eq. (5) and as we will show below, both the Kitaev interaction ($K$) and SIA ($A_{zz}$)  play an important role on the overall magnetic anisotropy of CrI$_3$ and CrGeTe$_3$. To demonstrate such fact, one can realize that Eq. (5) involves $(S_{X}S_{Y} + S_{Y}S_{Z} + S_{Z}S_{X})$, which adopts (i) its maximum when $S_{X} = S_{Y} = S_{Z} = \frac{\sqrt{3}}{2}$, which corresponds to spins being aligned along the out-of-plane z-direction; {\it versus} (ii) a minimum when $S_{X} + S_{Y} + S_{Z} = 0$, that is when spins are lying within the x-y plane. The sign and value of the $bK + \frac{2}{3}A_{zz}$ coefficient appearing in front of $(S_{X}S_{Y} + S_{Y}S_{Z} + S_{Z}S_{X})$ in Eq. (5) should therefore determine the magnetic anisotropy:  a negative $bK + \frac{2}{3}A_{zz}$  favors an easy axis along the out-of-plane direction while a positive $bK + \frac{2}{3}A_{zz}$ will encourage spins to lie within the x-y plane. To characterize the strength of such anisotropy between the out-of-plane direction and the x-y plane, we also computed  the energy difference, $\Delta{\mathcal{E}}$, between the energy of the state having a fully out-of-plane magnetization and the averaged energy of states having in-plane magnetization.

In the case of CrI$_3$, the $b$ parameter  is  numerically found to be negative. Together with the positive $K$ and negative $A_{zz}$ from Table 1, both $bK$ and $\frac{2}{3}A_{zz}$ are thus negative. They are determined to be -0.16 meV and -0.17 meV, respectively, as shown in Table 2. Such negative values indicate that both Kitaev interaction and SIA lead to an out-of-plane easy axis, which is further confirmed by the negative value of -1.11 meV/f.u for $\Delta{\mathcal{E}}$, as calculated from Eq. (5). Such value is not only consistent with the result of -0.82 meV/f.u. obtained from DFT calculations (confirming once again the validity and accuracy of our rather simple Eqs. (4) and (5)) but also explains the previously determined Ising behavior of CrI$_3$ favoring the out-of-plane direction for the magnetization\cite{huang2017layer,samarth2017condensed,griffiths1964peierls}.

In the case of CrGeTe$_3$, the  $b$ parameter is also found to be negative and leads to $bK$ adopting a  negative $-0.17$ meV value that is  similar to the one of CrI$_3$.  On the other hand,  $\frac{2}{3}A_{zz}$ is positive and yields $\simeq$ 0.17 meV, which therefore results in a nearly vanishing  $bK$+$\frac{2}{3}A_{zz}$ and thus to a $\Delta{\mathcal{E}}$ being nearly zero -- that is, -0.01 meV/f.u. according to Eq. (5), which also compares well with the result of 0.02 meV/f.u. directly obtained from DFT (the difference in sign between the energies from Eq. (5) and DFT can be overlooked since both calculations provide  vanishing $\Delta{\mathcal{E}}$). Such nearly zero value for $bK$+$\frac{2}{3}A_{zz}$ therefore implies that Eq. (5) predicts that CrGeTe$_3$ is basically isotropic in the whole three-dimensional space, that is any spatial direction of the magnetization (in-plane, out-of-plane or even combination between in-plane and out-of-plane components) should provide similar magnetic energy. Such finding is fully consistent with  the observed Heisenberg behavior of CrGeTe$_3$\cite{samarth2017condensed,xing2017electric,mermin1966absence}.
Note that, although the isotropic Heisenberg behavior of CrGeTe3 is confirmed by both previous experiments and the present computational work, the anisotropy (Kitaev interaction and SIA) still play a crucial important role in stabilization of the long-range magnetic ground state, as indicated by Mermin and Wagner¡¯s theorem\cite{mermin1966absence}.

\begin{figure}
\includegraphics[width=14cm]{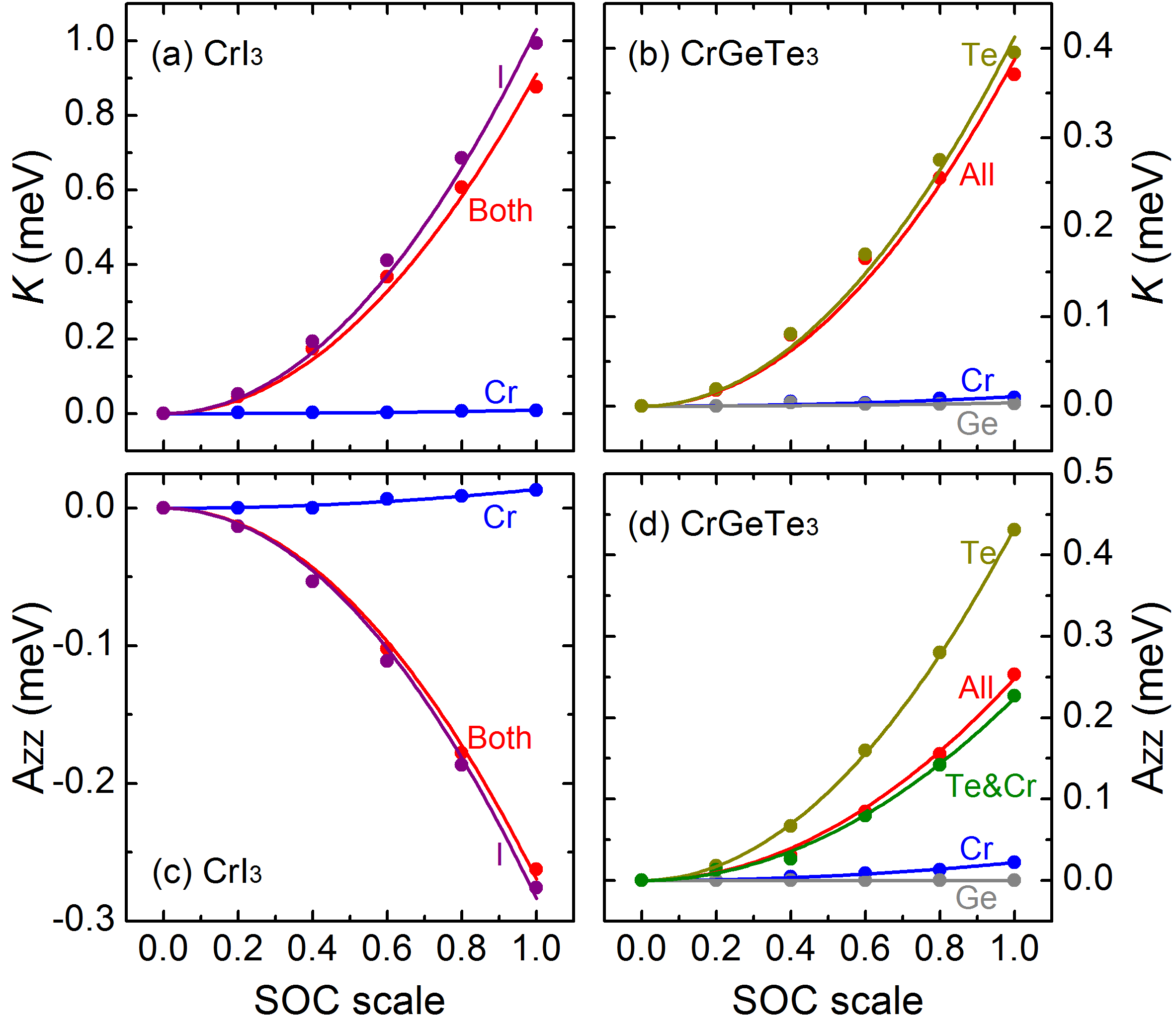}\centering%
\caption{Atomic dependencies of the Kitaev parameter $K$ and SIA coefficient $A_{zz}$. Panels {\bf a} and {\bf b} show the dependency of $K$ of CrI$_3$ and CrGeTe$_3$, respectively, as a function of the SOC strength. Panels {\bf c} and {\bf d} display the dependency of $A_{zz}$ for CrI$_3$ and CrGeTe$_3$, respectively, as a function of this SOC strength. Note that the value of 1.0 (0, respectively) for this SOC strength corresponds to the actual strength (no SOC, respectively) of the considered element.}
\end{figure}

It is also worthwhile to realize that both the Kitaev parameter $K$ and the SIA coefficient $A_{zz}$ originate from spin-orbit coupling. One may wonder what specific ions contribute to such coefficients via SOC. To address such issue, Figs. 2a and b display the atomically resolved contribution of the $K$ parameter as a function of the SOC strength in CrI$_3$ and CrGeTe$_3$, respectively. Such parameter mainly arises from the SOC of the heavy ligands (namely, I of CrI$_3$ or Te of CrGeTe$_3$) in a quadratic way, while the SOC of Cr has almost  no effect on $K$ (the effect from SOC of Ge on $K$ of CrGeTe$_3$ is also negligible).
Such predictions for CrI$_3$ and CrGeTe$_3$ are consistent with the results of Ref.\cite{lado2017origin} that anisotropy of CrI$_3$ mainly arises from the SOC of I ligand.
This can be understood by the fact that the super exchange between nearest neighbor Cr sites is mostly mediated by these ligands. To further confirm such results, we  developed a tight-binding model (see details in SM\cite{Note1}) that contains only two Cr ions and two bridging ligands, which form the $x'y'$ plane (see Fig. 3a for details). Such model confirms that $K>$ 0 and $K \varpropto \lambda^2$, where $\lambda$ is the SOC strength of ligands.

\begin{figure}
  \centering
  \includegraphics[width=16cm]{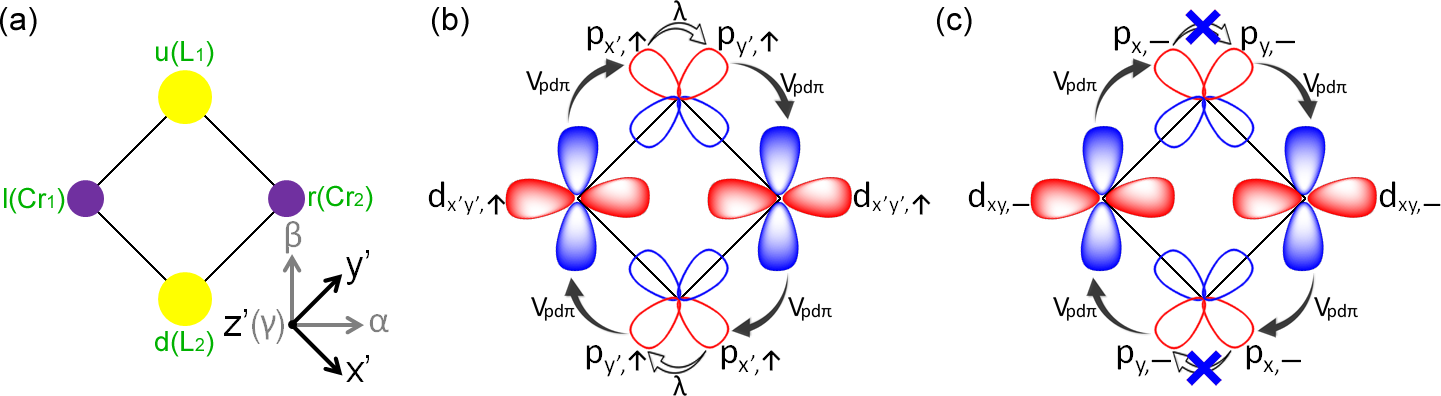}\centering
  \caption{Panel {\bf a}, the \{x'y'z'\} coordinate system and the configuration of the considered Cr$_2$L$_2$ cluster in the tight binding model, where L is a ligand ion. Panel {\bf b}, schematization of the extra hopping path related to spins being along the $z'$ ($\uparrow$) direction. Panel {\bf c}, the forbidden hopping paths related to spins lying in the $x'y'$ ({\bf --}) plane, respectively.}
\end{figure}

We now work on understanding the analytical results that $K>0$ and $K\varpropto\lambda^2$. If SOC is not considered, the magnetic coupling is isotropic with the strength of $J$. When the SOC of ligands is included, an extra hopping path emerges for spins that are along the $z'$ direction, as shown in Figs. 3a and b, since
$\langle p_{x'\uparrow}|\mathbf{L}\cdot\mathbf{S}|p_{y'\uparrow}\rangle \neq 0 $.
Such an extra path provides an additional energy term $K$ to the magnetic coupling, as $J_{z'z'}=J+K=\dfrac{1}{2S^2}(E_{FM,z'} - E_{AFM,z'})$.

In the antiferromagnetic case, the two Cr have opposite spins. The electrons can hop from the occupied $t_{2g}$ orbitals of one Cr to the unoccupied $t_{2g}$ orbitals of another Cr, which can lower the $E_{AFM,z'}$ by the amount of $K\varpropto\lambda^2$. The form of $\lambda^2$ can be understood as that the whole hopping procedure includes two times of ligands' SOC effects, as shown in Fig. 3b.
On the other hand, in the ferromagnetic case, $t_{2g}$ orbitals of both Cr are occupied with electrons having the same spin direction. In such case, although the aforementioned extra hopping path still exists, it can not lead to the energy lowering of $E_{FM,z'}$.
In contrast, there is no such extra hopping path when spins lie in the Cr$_2$L$_2$ ($x'y'$) plane (denoted with the ``--'' mark), since
$ \langle p_{x'-}|\mathbf{L}\cdot\mathbf{S}|p_{y'-}\rangle = 0 $.
Such effects are further illustrated in Fig. 3c. As a result, the total effective $J_{z'z'} = J + K$ is larger than that $J_{x'x'} = J_{y'y'} = J$, i.e., $K>$ 0.
The findings reported here thus demonstrate, for the first time, that  (i) the Kitaev interaction not only exists in 4$d$ or 5$d$ transition metal insulators, but also can occur in 3$d$ systems; and (ii) the SOC of the ligands can  play a crucial  role on that interaction. Such findings can  facilitate the ongoing efforts to realize Kitaev-type interactions in 3$d$ systems\cite{liu2018pseudospin,sano2018kitaev}.

Regarding SIA, Fig. 2c shows that  the SOC of I is basically responsible for the negative $A_{zz}$ of CrI$_3$, and that Cr does not significantly contribute to such parameter. Such finding of a ligand-induced SIA is also novel, since SIA is typically believed to arise from the transition metal ion\cite{xiang2008origin,hu2014giant}. Similarly and as indicated by Fig. 2d, the Te ligand in CrGeTe$_3$ produces a rather strong (and positive, in that case)  $A_{zz}$, which is even twice as large as the one resulting from the SOC of all considered ions. It is in fact the combination of SOC from both Te and Cr that provides a value of $A_{zz}$ that is close to the total one in CrGeTe$_3$ (note that Ge does not significantely contribute to this total SIA).

\noindent
{\bf DISCUSSION}\\
{\it Explanations on magnetic behaviors of related 2D systems.} Another 2D ferromagnetic system, CrBr$_3$, was also studied. It is numerically found (not shown here) that, as similar to CrI$_3$, both the Kitaev interaction and SIA favor an out-of-plane easy axis in this material. Such finding is consistent with measurements determining that the net anisotropy of CrBr$_3$ is out-of-plane\cite{mcguire2017magnetic}.
Interestingly, Eq. (5) of the manuscript can also be useful to shed some light into controversial issues, such as which magnetic model is more pertinent to CrSiTe$_3$. As a matter of fact, early neutron work suggested an Ising-like model\cite{carteaux19952d} for this system, while recent measurements argue between an Heisenberg-like model\cite{williams2015magnetic} and an Ising model coupled with long-range interaction\cite{liu2016critical}. CrSiTe$_3$ shares similarities with  CrGeTe$_3$ in the sense that it has a negative $bK$ associated with  Kitaev interaction, as well as a positive $\frac{2}{3}A_{zz}$ induced by SIA. However, the former is equal to $-$0.21 meV and is larger in magnitude than the latter (that is equal to +0.11 meV) in CrSiTe$_3$, therefore leading to a less negative $bK$+$\frac{2}{3}A_{zz}$ of $-$0.07 meV and thus slightly tipping the balance towards out-of-plane magnetization (as confirmed by our DFT results providing  $\Delta{\mathcal{E}}$=-0.10 meV for the difference between the energy for an out-of-plane  magnetization and the averaged energy of ferromagnetic states having in-plane magnetization). One can thus propose that the correct magnetic model for CrSiTe$_3$ should be related to a slight perturbation of the Ising model, in order to account for possible (weaker) in-plane components of the magnetization in addition to (stronger) out-of-plane ones.

\noindent
{\it Applications of the present general model and the XXZ model.} The present general model (Eqs. 1, 4 and 5) adopts the most generalized form of the $\mathcal{J}$ and $\mathcal{A}$ matrices, which can capture the microscopic details of different anisotropy. Such model is powerful, as it explains the origin of Kitaev interaction, as well as the competition and collaboration between Kitaev interaction and SIA. It also allows for antisymmetric exchange coupling, i.e. the so called Dzyaloshinskii-Moriya interaction, when the inversion centers between Cr-Cr pair are somehow removed. In contrast, the XXZ model is more macroscopic in nature, since it starts from the overall effects of the frustration among Cr-Cr pairs. Nevertheless, this XXZ model can still somehow describe the competition and collaboration between Kitaev interaction (since $\mathcal{J}_x$ and $\mathcal{J}_z$ are different from each other) and SIA. As a result,  the XXZ model can be technically applied to both CrI$_3$ and CrGeTe$_3$, as well as to the aforementioned related systems, but, as documented  in section 4 of the Supplemental Materials, one really has to include SIA there to be more accurate.

\noindent
{\it Implication of potential Kitaev-type quantum spin liquid.} The predicted presence of Kitaev interaction in CrI$_3$ and CrGeTe$_3$ systems hints towards the possibility of realizing quantum spin liquid state in 3$d$ systems. As a matter of fact, additional calculations we performed (not shown here) indicate that varying in-plane strain can make the isotropic exchange coupling vanishing while the Kitaev interaction remains finite, which is promising to realize quantum spin liquids state in CrI$_3$ and CrGeTe$_3$ systems.
Moreover, our predictions also provide another way to enhance Kitaev interaction in the ``traditional'' 4$d$ and 5$d$ systems, that is, for example, to substitute the light Cl ligand with the heavier I ion in RuCl$_3$ compound. The hybrid source of SOC to produce strong Kitaev interaction should result in interesting physics and phenomena in related systems.

We hope that our first-principles calculations and concomitant development of a simple insightful Hamiltonian (see Eqs. (4) and (5)), along with a tight-binding model, deepens the understanding of magnetic anisotropy in low-dimensional systems.  The decomposition of the total magnetic anisotropy into Kitaev and SIA effects further sheds light into the behaviors of other related systems, such as CrBr$_3$ and CrSiTe$_3$ (see SM\cite{Note1}), therefore further demonstrating its relevance and importance.

\begin{methods}
See supplementary materials\cite{Note1} for additional details about the methods used here, as well as further information about our predictions.

All data generated or analysed during this study are included in this published article (and its supplementary information files).

\end{methods}


\begin{addendum}
 \item We thank Zhenglu Li and Jianfeng Wang for useful discussion. This work is supported by the Office of Basic Energy Sciences under contract ER-46612. H.X. is also supported by NSFC (11374056), the Special Funds for Major State Basic Research (2015CB921700), Program for Professor of Special Appointment (Eastern Scholar), Qing Nian Ba Jian Program, and Fok Ying Tung Education Foundation. The Arkansas High Performance Computing Center (AHPCC) is also acknowledged.
 \item[Author contributions] C.X. and J.F contributed equally to this work. H.X and C.X. conceived the idea. H.X. and L.B. supervised this work. C.X. performed the DFT calculations and related analysis, J.F. worked on the tight-binding model. All authors contribute to the discussion of the results and the writing of the manuscript.
 \item[Competing Interests] The Authors declare no Competing Financial or Non-Financial Interest.
 \item[Correspondence] Correspondence should be addressed to
        H.X. (email: hxiang@fudan.edu.cn) and L.B. (email: laurent@uark.edu).
\end{addendum}



\end{document}